\documentclass[preprint,nofootinbib]{revtex4-1}
\usepackage{amssymb,amsmath,amsfonts,mathrsfs,graphicx,xcolor,units,yfonts}
\usepackage{hyperref}
\usepackage[capitalise]{cleveref}
\usepackage{enumitem}

\hypersetup{
  colorlinks=true,        
  linkcolor=blue,         
  citecolor=blue,       
  urlcolor=black  }

\newcommand{\dmunu}{_{\mu\nu}}

\newcommand{\ivc}[1]{{\color{black} #1}}
\newcommand{\uzc}[1]{{\color{black} #1}}

\begin{document}
\title{Deformation of horizons during a 
  merger}
\author{Uzair Hussain}
\email[]{uh1681@mun.ca}
\author{Ivan Booth}
\email[]{ibooth@mun.ca}
\affiliation{Department of Mathematics and Statistics,   Memorial University of Newfoundland  St John's NL A1C 4P5, Canada}
\date{\today}

\begin{abstract}
We model an extreme mass ratio merger (EMR) as a point particle radially plunging into a large Schwarzschild black hole.  
We assume that the mass of the point particle, $\mu$, is much smaller than the black hole mass $M$. Under this assumption we can employ the Zerilli formalism modified to include a source term which arises from the energy-momentum tensor of the small object. We solve the Zerilli equation by numerically evolving initial data.  Then, we ray trace the null geodesics of the event horizon from after the merger backward in time to extract  the geometry of the perturbed event horizon. Further, we take advantage of the axisymmetry of the setup to locate the apparent horizon and study its geometry.
\end{abstract}

\maketitle
\section{Introduction}
From the gravitational radiation observed by LIGO's interferometers we now have strong evidence pointing towards the existence of binary black hole systems and their inevitable mergers \cite{Abbott:2016}. To predict the waveforms of the radiation emitted by these mergers complex simulations are executed on super computers using state-of-art algorithms. These simulations can accommodate a wide range of the possible parameters of the binary system, for e.g., initial spins, spin orientations, mass ratios, initial separation and velocities, etc. 

In a very specific type of merger the calculations can be handled by much simpler methods. This happens when we consider the head-on merger of two non-spinning black holes which have an extreme mass ratio (EMR): that is the mass, $\mu$, of one of the black holes is much smaller than the mass, $M$, of the other, i.e. $\mu/M \rightarrow 0$. Then focusing on the large black hole, one can model the smaller black hole as a point particle with mass $\mu$. Hence, the problem is reduced to the radial in-fall of a point particle into a Schwarzschild black hole.

This approach to the problem was pioneered by Zerilli \cite{Zerilli:1971} and has since been studied extensively \cite{Martel:2001,Spallicci:2010,Lousto:1997}. At the heart of the solution is the assumption that the gravitational field produced by the particle drops off rapidly and so only perturbs the surrounding geometry. Further, there is no back-reaction on the particle and thus it follows a geodesic in the spacetime of the large Schwarzschild black hole. Under these assumptions Einstein's equations can be linearized. 

Here we will study how, in this perturbed spacetime, the event and apparent horizons of the large black hole are deformed by the gravitational effects of the stress-energy tensor of the point particle. For the deformation of the event horizon, a prominent role is played by caustics: points where null rays in the vicinity of the large black hole's horizon cross each other and join the horizon. Recently, similar results have been shown in \cite{Emparan:2016}. There the strategy was to take the limit $M\rightarrow \infty$ and model the large black hole as a Rindler-type planar horizon accelerating towards the small black hole, which is assumed to be Schwarzschild. Then as the small black hole approaches, the null generators of the planar horizon are deflected towards the small black hole due to its Schwarzschild geometry and as such the deformation of the planar event horizon and the smaller hole's event horizon can be computed. Another study that shows a similar deformation of the event horizon is \cite{Hamerly:2010}, where the approach is the one we take: approximating the small black hole as a point particle and solving for the perturbed geometry. 
In that paper the influence of the particle is treated as an impulse in the frequency domain of the perturbations.

By contrast, in this study we will do a direct numerical evolution, in the time domain, of Brill-Linquist type initial data for binary black holes. 
As in previous studies, we find that the event horizon of the large black hole ``reaches out'' in a cusp towards infalling particle. By contrast, for the 
apparent horizon we find that it appears to recede away from the approaching particle. While this is at first counter-intuitive, 
a little consideration shows that this isn't so strange. Intuitively, as the particle gets closer there are null geodesics inside the black hole, close to the horizon, which get pulled outwards by the particle's gravitational field. Hence the zero expansion surface moves inwards. Similar behaviour has 
been observed examining the marginally outer trapped surfaces (MOTS) in slices of initial data for arbitrary mass ratio head-on collisions 
\cite{Mosta:2015}.

The paper is organized as follows. Section \ref{Pert} reviews the perturbation formalism for the radial in-fall of a point particle, Section \ref{numerical} outlines the numerical method used to evolve the initial data, Section \ref{deform} shows how the event and apparent horizons are deformed, and finally we conclude with Section \ref{conclusion}. \ivc{Appendix \ref{l1pert} examines gauge choices  for the $\l=1$ modes, Appendix \ref{ecviolations} examines the physical effects of one of the approximations that 
we make in the course of our calculations and Appendix \ref{convergence} tests numerical convergence of our schemes. }

\section{Perturbations by radially in-falling point particle}\label{Pert}

We assume the particle creates a small disturbance in the spacetime which can be modelled by standard perturbation theory methods. We use the approach of Zerilli  \cite{Zerilli:1971}  with the Moncrief wavefunction \cite{Moncrief:1974} and express the metric as $\tilde{g} \dmunu = g \dmunu + p \dmunu$, where $g \dmunu$ is the Schwarzschild metric and $p \dmunu$ is  the perturbation. Given the spherical symmetry of the Schwarzchild spacetime, $p \dmunu$ can be expanded in terms of odd and even modes with indices $l$ and $m$ arising from the spherical harmonics. If we assume the particle falls along the $z$-axis, then we can see that the problem is axisymmetric and only even modes are excited \cite{Martel:2001}. We write the perturbation as,
\begin{align}
p^{l0} \dmunu dx^\mu dx^\nu = Y^{l0}(\theta) \left( f H_0^{l0} dt^2 + 2 H_1^{l0} dt dr + \frac{1}{f} H_2^{l0} dr^2 +r^2 K^{l0} d \Omega^2 \right)
\end{align} 
where we have taken $m=0$ again because of axisymmetry, $f(r)=1-2M/r$ and the $H_0, H_1, H_2$ and $K$ are functions of $r$ and $t$, we omit 
the $l0$ where there is no chance of confusion. We work in the Regge-Wheeler gauge and  Schwarzschild coordinates.

Then invoking Einstein's equations up to linear order in $p\dmunu$, we find that the equations can be decoupled into one inhomogeneous wave equation:
\begin{align}\label{master}
\left[ -\frac{ \partial^2 }{\partial t^2} + \frac{ \partial^2}{\partial r^{*2}} - V(r)\right] \psi(r,t) = S(r,t)
\end{align}
where $r^*= r+2M \text{ln} (r/2M-1)$ is the tortoise coordinate,
\begin{align}
\psi(r,t)= \frac{r}{\lambda+1} \left \{ K + \frac{f}{\Lambda} \left[H_2-r\frac{\partial}{\partial r} K \right ] \right\} 
\end{align}
is the Zerilli-Moncrief function,
\begin{align}
V(r)=\frac{2 f}{r^2 \lambda^2} \left[ \lambda^2 (\Lambda+1)+\frac{9M^2}{r^2} \left(\Lambda - \frac{2M}{r}\right)\right]
\end{align}
is the even potential with $\lambda=(l+2)(l-1)/2$, $\Lambda=\lambda+\frac{3M}{r}$, and
\begin{align}\label{source}
S(r,t)=& \frac{2}{l(l+1)\Lambda} \Bigg\{ r^2 f \left[ f^2 \frac{\partial }{\partial r} Q^{tt} - \frac{\partial }{\partial r} Q^{rr} \right] \\ &- \frac{f^2}{\Lambda r} \left[ \lambda(\lambda-1)r^2+ (4\lambda -9) M r+15 M^2 \right] Q^{tt}+ r(\Lambda-f) Q^{rr} \Bigg\} 
\end{align}
is the source term for radial infall. The tensor $Q^{ab}$ where $a,b$ run over $r,t$ is given by,
\begin{equation}\label{Qab}
Q^{ab}= 8\pi \int T^{ab} Y^{*}(\theta,\phi) d\Omega
\end{equation}
\uzc{where $Y^{*}(\theta,\phi)$ is the complex conjugate of a spherical harmonic (notice the omitted $l$)}, $T^{ab}=\mu \int u^a u^b \delta^4 (x^\nu - x_p^\nu (\tau)) \sqrt{-g} d \tau$, is the stress tensor of the particle, $\mu$ is the mass of the particle, $\tau$ is the proper time along the particle's trajectory, $x_p(\tau)$, and the four-velocity of the particle is,
\begin{equation}\label{fourvector}
u^\nu = (\tilde{E}/f, -(\tilde{E}^2-f)^{1/2},0,0)
\end{equation}
where $\tilde{E}$ is the conserved energy. Thus
\begin{align}\label{qtensor}
Q^{ab} =\mu \frac{8 \pi}{r^2} \frac{u^a u^b}{u^t} \delta(r-r_p(t)) Y^*(\theta(t), \phi(t)) \, . 
\end{align}
\uzc{Here $Y^*(\theta(t), \phi(t))$ means that the spherical harmonic is evaluated along the trajectory of the particle, since the particle moves radially and its angular coordinates do not change}, this is a constant. Inserting (\ref{qtensor}) into (\ref{source}) 
\begin{equation}
S(r,t)=16 \pi \frac{\sqrt{(2l+1)/4\pi}}{l(l+1)} \frac{\mu}{\tilde{E}} \frac{f^3}{\Lambda} \left\{ \delta'(r-r_p(t)) - \left[ \frac{(\lambda+1)r -3M}{r^2f} - \frac{6M\tilde{E}^2}{\Lambda r^2 f}\right] \delta(r-r_p(t)) \right\}
\end{equation}
where $r_p(t)$ is the trajectory of the particle and $Y^{*l0}(0)=\sqrt{(2l+1)/4\pi}$. The equation for the trajectory starting from rest is,
\begin{equation}\label{traj}
\frac{t}{2M}=\tilde{E} \sqrt{1-\frac{r}{r_0}} \sqrt{\frac{r}{2M}}  \frac{r_0}{2M}  + 2 \text{arctanh} \left(\frac{\sqrt{\frac{2M}{r}-\frac{2M}{r_0}}}{\tilde{E}}\right) + \tilde{E} \left(1+\frac{4M}{r_0}\right) \left(\frac{r_0}{2M}\right)^{3/2} \text{arctan} \left( \sqrt{\frac{r_0}{r}-1} \right)
\end{equation}
where $r_0$ is the initial position of the particle and $\tilde{E}=\sqrt{1-\frac{2M}{r_0}}$.
\section{Numerical method}\label{numerical}
We use the numerical method of Lousto and Price \cite{Lousto:1997} with the modification by Martel and Poisson \cite{Martel:2001}. The method is a finite difference algorithm on a staggered grid with a step size $\Delta=\Delta r^*/4M=\Delta t/2M\approx 0.009$ over a domain bounded by two null hypersurfaces, $\mathcal{H}_1$ and $\mathcal{H}_2$,  and one spacelike surface $\Sigma_0$. $\mathcal{H}_1$ approximates the event horizon and is given by, $u=t_f-r^*_p(t_f)$ where $t_f$ is determined by (\ref{traj}) with $r_p(t_f)/2M=1.00001$. $\mathcal{H}_2$ is a null hypersurface which approximates future null infinity and is chosen to be, $v/2M=(t+r^*)/2M\approx 1500$. Finally $\Sigma_0$ is the $t=0$ surface which is the moment of time symmetry for the initial data.

\begin{figure}[h]
\includegraphics[scale=1.0]{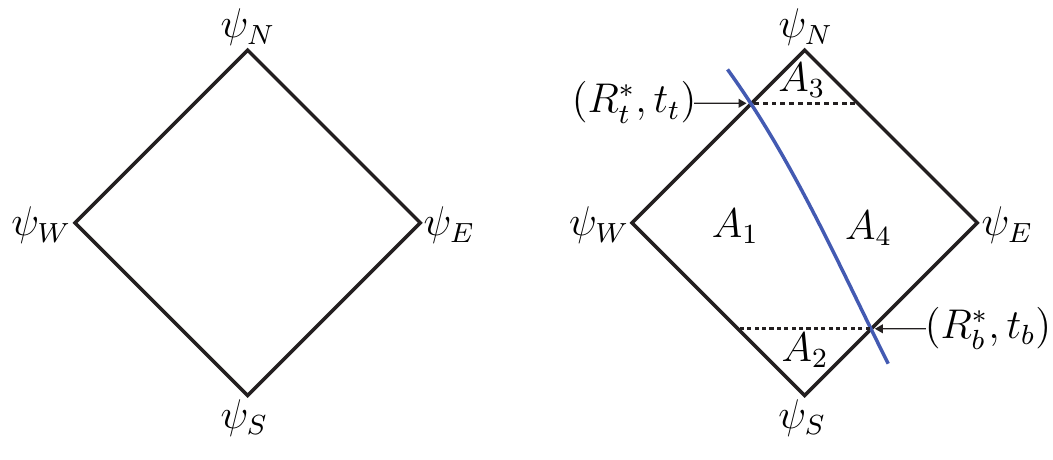}
\caption{These are the two types of cells that arise when discretizing the domain. The left one (Type I) does not contain the particle and hence we do not need to integrate over a source term. The right one (Type II) contains the particle and hence we have to integrate over the delta function source.}\label{cell}
\end{figure}

After discretizing the domain we encounter two kinds of cells, as shown in FIG. \ref{cell}; ones that do not contain the particle  trajectory (Type I) and ones that  that do (Type II). For Type I cells we do not have a contribution from the source term, $S(r,t)$. So we may discretize (\ref{master}) with the following formula,
\begin{equation}
\psi_N=-\psi_S+\left(\psi_W+\psi_E\right)\left(1-\frac{\Delta^2}{2}V_S\right)
\end{equation}
where $S=(r^*_S,t_S)$ is the base point, $W=(r^*_S-\Delta,t_S+\Delta)$, $E=(r^*_S+\Delta,t_S+\Delta)$ and $N=(r^*_S,t_S+2\Delta)$. This scheme is accurate to $\mathcal{O}(\Delta^4)$. 

For Type II cells there is a contribution from the source term and (\ref{master}) now takes the form\footnote{There is a typo in \cite{Martel:2001} for equations (\ref{disc1}) and (\ref{disc2}), see \cite{Spallicci:2010}.}
\begin{align}\label{disc1}
\psi_N=&-\psi_S\left[1+\frac{V_S}{4}(A_2-A_3)\right]+\psi_E\left[1-\frac{V_S}{4}(A_3+A_4)\right] \nonumber\\ &+\psi_W\left[1-\frac{V_S}{4}(A_1+A_3)\right]-\frac{1}{4}\left(1-\frac{V_b}{4}A_3\right) \iint dA S(r,t)
\end{align}
where the $A_i$'s are the areas illustrated in right panel of FIG. \ref{cell} and $\iint dA S(r,t)$ is the integration of the source term over the cell,
\begin{align}\label{disc2}
\iint dA S(r,t)=& -\kappa\int_{t_b}^{t_t} dt \frac{f(t)}{\Lambda(t)^2} \frac{1}{r_p(t)}\left[ \frac{6M}{r_p(t)} (1-\tilde{E}^2)+\lambda (\lambda+1)-\frac{3M^2}{r_p(t)^2}+4\lambda \frac{M}{r_p(t)}\right] \nonumber\\ & \pm \kappa \left\{ \frac{f(t_b)}{\Lambda(t_b)} [1\mp  \dot{r}^*_p(t_b) ]^{-1} +\frac{f(t_t)}{\Lambda(t_t)} [1\pm  \dot{r}^*_p(t_t) ]^{-1}\right\}
\end{align}
where, $l(l+1) \tilde{E} \kappa = 16 \pi \mu \sqrt{(2l+1/(4\pi)}$,  $\dot{r}^*_p(t)= -\sqrt{\tilde{E}^2-f(t)}/\tilde{E}$, $f(t)=f(r_p(t))$, $\Lambda(t)=\Lambda(r_p(t))$, $t_b$ $(t_t)$ is the time when the particle enters (leaves) the cell. Further, in the expression on the second line, the upper (lower) sign for the first term (function of $t_b$) is used when the particle enters the cell on the right (left) of $r^*_S$. For the second term (function of $t_t$) the upper (lower) sign is used when the particle leaves the cell on the right (left) of $r^*_S$.
\begin{figure}[h]
\includegraphics[scale=0.8]{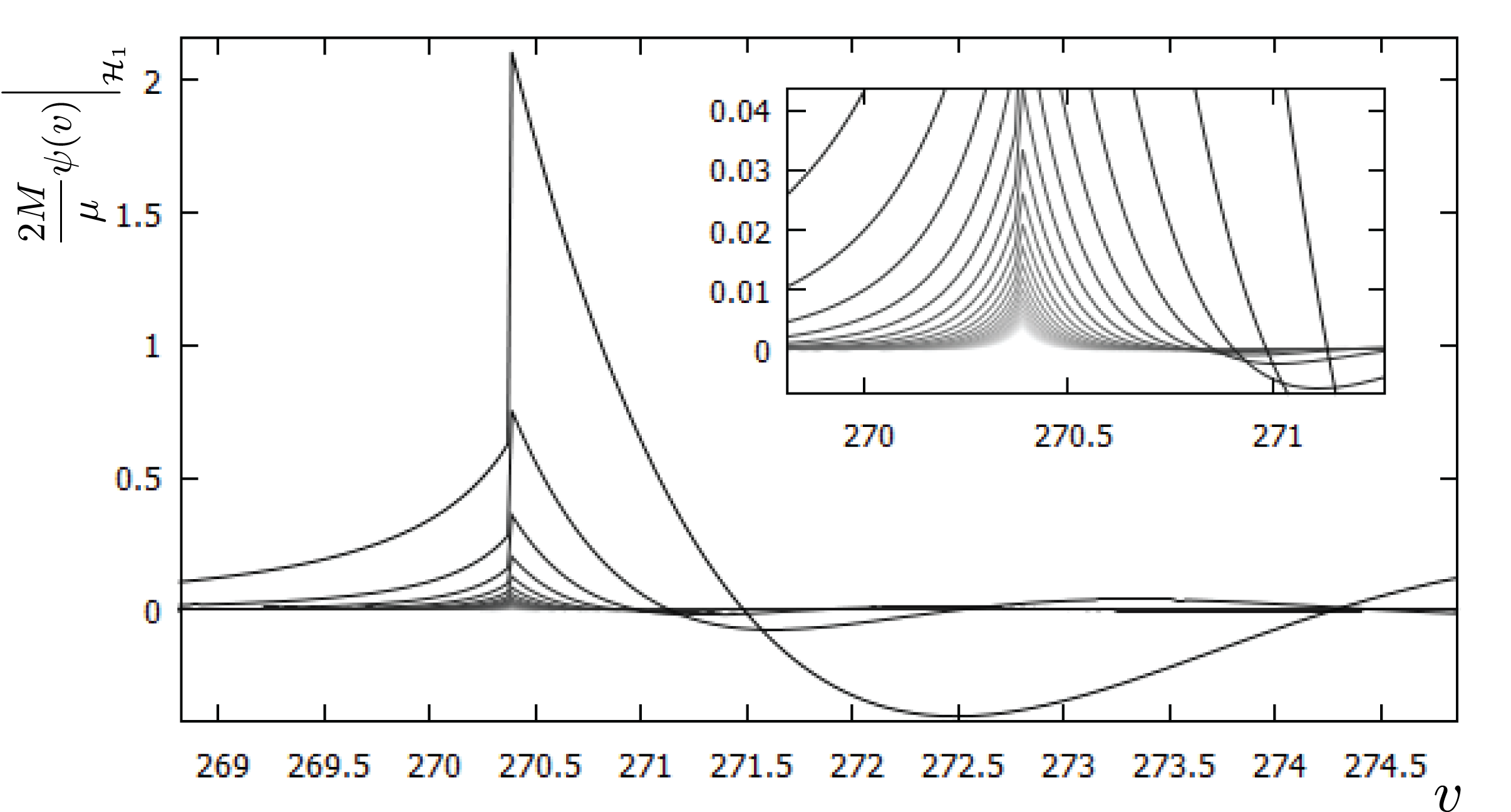}
\caption{Waveforms, $\psi(v)$ extracted at $\mathcal{H}_1$ for $l=2$ to $l=25$. The lower the amplitude the higher the value of $l$. The inset is zoomed in at the time when the particle meets the event horizon.} \label{waveforms}
\end{figure}
The initial data at $t=0$ is constructed from the Brill-Lindquist \cite{Brill:1963} solution and is  therefore conformally flat. It has been analysed in detail in \cite{Martel:2001,Lousto:1997, Lousto:1996}, and is given by,
\begin{equation}\label{initial}
H_2=K=2\mu \frac{\sqrt{4\pi/(2l+1)}}{(1+M/2\bar{r}_0)(1+M/2\bar{r})} \frac{\bar{r}_{<}^l}{\bar{r}_{>}^{l+1}}
\end{equation}
where $\bar{r}=r(1+\sqrt{f})^2/4$ is the isotropic radius, and $\bar{r}_< (\bar{r}_>)$ is the smaller (greater) of $\bar{r}$ and $\bar{r}_0$. Note that with this choice of initial data, when we  evolve from the $\Sigma_0$ surface, we have not specified what the $\psi_S$ points will be. We circumvent this problem by doing a Taylor expansion, $\psi(-\Delta,r^*)=\psi(\Delta,r^*)-2\Delta \partial_t \psi(t_0,r^*)+\mathcal{O}(\Delta^3)$, since the data is time symmetric we can take $ \psi(-\Delta,r^*)=\psi(\Delta,r^*)$. Since we will be interested in the horizon deformations, we will extract $\psi$ on the $\mathcal{H}_1$ surface. FIG. \ref{waveforms} shows the waveforms, $\psi(v)$, for $l=2$ to $l=25$ extracted on the surface $\mathcal{H}_1$ as a function of the ingoing coordinate $v$ with $r_0/2M=30$. \uzc{This choice of $r_0$ makes $\tilde{E}\approx 1$.}  Notice that for large $l$ we get a smaller amplitude. Hence, we introduce a cutoff, $l_{\text{cut}}$, at $l_{\text{cut}}=25$. 
The (small) physical effect of this approximation is discussed in  Appendix \ref{ecviolations}.

\uzc{In Appendix \ref{convergence} we have shown how the code converges, where the method used to show this is the same as in \cite{Martel:2001}. Further, as a small test of our code we present the results of a simulation with a set of parameters chosen from \cite{Martel:2001}. Specifically, we extract the $l=2$ mode for $\psi$ at the surface that approximates null infinity, $\mathcal{H}_2$. We do this for a particle which starts from rest with a starting position of $r_0/2M=40$. In \cite{Martel:2001} the result with these parameters are presented in FIG. 3 (first plot, with $\alpha=1$), here it is presented in FIG. \ref{comparetopoisson}. By visual inspection  these curves are virtually identical.}

\section{Horizons deformation}\label{deform}
\subsection{Event Horizon}
To find the deformation of the event horizon we use a method similar to the one employed in \cite{Hamerly:2010}, which calculates how each null generator of the horizon gets perturbed from its usual Schwarzschild position. For another method which locates the whole event horizon as a null surface, see \cite{Anninos:1994}. Consider the null generators of the event horizon of the static Schwarzschild spacetime in Kruskal-Szekeres coordinates, \uzc{in terms of $u=t-r*$ and $v=t+r*$, these coordinates are defined as,
\begin{equation}
U=-e^{-\kappa u}\hspace{0.5 cm}\text{and}\hspace{0.5cm}V=e^{\kappa v}.
\end{equation}
The geodesic equation for the generators is given by,
\begin{align}\label{geodesic}
\frac{d^2 X^\mu}{dV^2}=-\hat{\Gamma}^\mu_{\sigma\gamma} \frac{d X^\sigma}{dV}  \frac{d X^\gamma}{dV}+g \frac{d X^\mu}{dV}
\end{align}
where $X^0=V(V),X^1=U(V)=0,X^3=\theta(V), X^4=\phi(V)$}, $\hat{\Gamma}^\mu_{\sigma\gamma}$ denotes the Christoffel symbols in the Kruskal-Szekeres coordinates and we have kept the non-affine parameter $g$; for the static case we have $g=0$. In our case  there is a perturbation present and so we add a perturbative term to each of the $X^i$ where $i$ runs over $U$ and $\theta$. We have then, $X^i\rightarrow X^i + \delta X^i$. Also, the Christoffel symbols will be perturbed, $\hat{\Gamma}^\mu_{\sigma\gamma} \rightarrow \hat{\Gamma}^\mu_{\sigma\gamma} + \delta \hat{\Gamma}^\mu_{\sigma\gamma}$. Inserting this \emph{ansatz} into (\ref{geodesic}) and appropriately evaluating on the horizon we get for $i=\theta$;
\begin{align}\label{deltathetav}
\frac{d^2 \delta \theta}{dV^2}=-\delta\hat{\Gamma}^\theta_{VV} 
\end{align}
for $\delta U$ we may use the null condition for the generators,
\begin{align}\label{deltauv}
 \frac{d \delta U}{dV}= e \kappa^2\delta g_{VV}
\end{align}
Switching to ingoing coordinates and expressing the perturbed quantities in terms of $\psi$,  we can write (\ref{deltathetav}) and (\ref{deltauv}) as, 
\begin{align}
\left( \frac{\partial}{\partial v} -\kappa \right) \frac{\partial \delta \theta^l}{\partial v}=\kappa \left( \frac{\partial }{\partial v} -\kappa \right) \frac{\partial \psi^l}{\partial v}  \partial_\theta Y^l
\end{align}
\begin{align}
\left( \frac{\partial }{\partial v} -\kappa \right) \delta r^l = -\frac{1}{2\kappa} \left(  \frac{\partial }{\partial v} -\kappa \right) \frac{\partial \psi^l}{\partial v}  Y^l
\end{align}
where $l$ is the mode number for the spherical harmonics. The general solution is given by,
\begin{equation}
\delta \theta^l = \kappa \psi^l \partial_\theta Y^{l0} + \frac{1}{\kappa} \theta_0^l e^{\kappa v} + \theta_1^l
\end{equation}
\begin{equation}
\delta r^l = -\frac{1}{4\kappa} \partial_v \psi^l Y^{l0}+ R_0^l e^{kv} 
\end{equation}
where $R_0^l,\theta_0$ and $\theta_1$ are constants. 

To find the horizon deformation we need to evolve the equations `backwards in time'. We start from the null generator of a static black hole at $U=0$ and $V\rightarrow \infty$ and integrate backwards to extract the deformation of the horizon. Hence, we impose the requirement that $\delta \theta$ and $\delta r$ vanish for $v \gg v_m $ where $v_m$ is time when the particle and black hole meet. Since at late times the perturbation vanishes we choose $R_0^l=\theta_0=\theta_1=0$. For $v=v_m$, note that there is a discontinuity in $\psi$ (cf. FIG. \ref{waveforms}). To get a  continuous deformation we choose:
\begin{align}
R_0^l&= -\frac{1}{4\kappa}\Delta \psi^l_{,v} e^{-k v} |_{v_{-}} Y^{l} \label{constant1} \\ \theta^l_0  &= \kappa \Delta \psi^l_{,v} \partial_{\theta} Y ^{l}  e^{-k v} |_{v_{-}} \label{constant2}\\
\theta_1^l&=(\kappa \Delta \psi^l - \Delta \psi^l_{,v})  \partial_{\theta} Y ^{l} \label{constant3}
\end{align}
where,
\begin{equation}
\Delta \psi^l = \psi^l|_{v_{+}} - \psi^l|_{v_{-}} \hspace{1cm} \Delta \psi^l_{,v} = \psi^l_{,v}|_{v_{+}} - \psi^l_{,v}|_{v_{-}}
\end{equation}  
where $v_{+(-)}$ are the limits from the right (left) of $v_m$.  

{
The effect of the monopole term, $l=0$, is calculated in the following manner: since the monopole term is isotropic, we use (\ref{deltauv}) for $v>v_m$ and arrive at,
\begin{equation}\label{monopole}
\left( \frac{\partial }{\partial v} -\kappa \right) \delta r^0 = -\frac{\mu}{2M}
\end{equation}
which has a general solution  $\delta r^0 = 2 \mu + C_0 e^{\kappa v}$, where $C_0$ is  a constant. For $v<v_m$ the right hand side of (\ref{monopole}) will be zero and as such we have $\delta r^0 = C_0 e^{\kappa v}$.  We fix $C_0=0$ for $v>v_m$ and for continuity at $v_m$ we fix $C_0= 2\mu e^{\kappa (v -v_m)}$ for $v<v_m$.} 

\ivc{The $l=1$ mode is a little more complicated. Intuitively it corresponds to linear motions of the black hole and translations of the coordinate system. Ideally one would like to pick the 
``centre-of-mass'' gauge for which the system is stationary at infinity. Unfortunately that gauge turns out to be unsuitable, with key components of the  metric diverging on the horizon\footnote{While 
this can be seen algebraically in the discussion of Appendix \ref{l1pert} as of now we don't have a good intuitive explanation for this result. We will return to this point in future work.}. Hence we instead 
perform an infinitesimal gauge transformation to go into the singular gauge \cite{detweiler2004low} in which the $l=1$ contribution to horizon deformations vanishes. These points are discussed
in more detail in  Appendix \ref{l1pert}. }

\begin{figure}[h!]
\centerline{\includegraphics[scale=0.70]{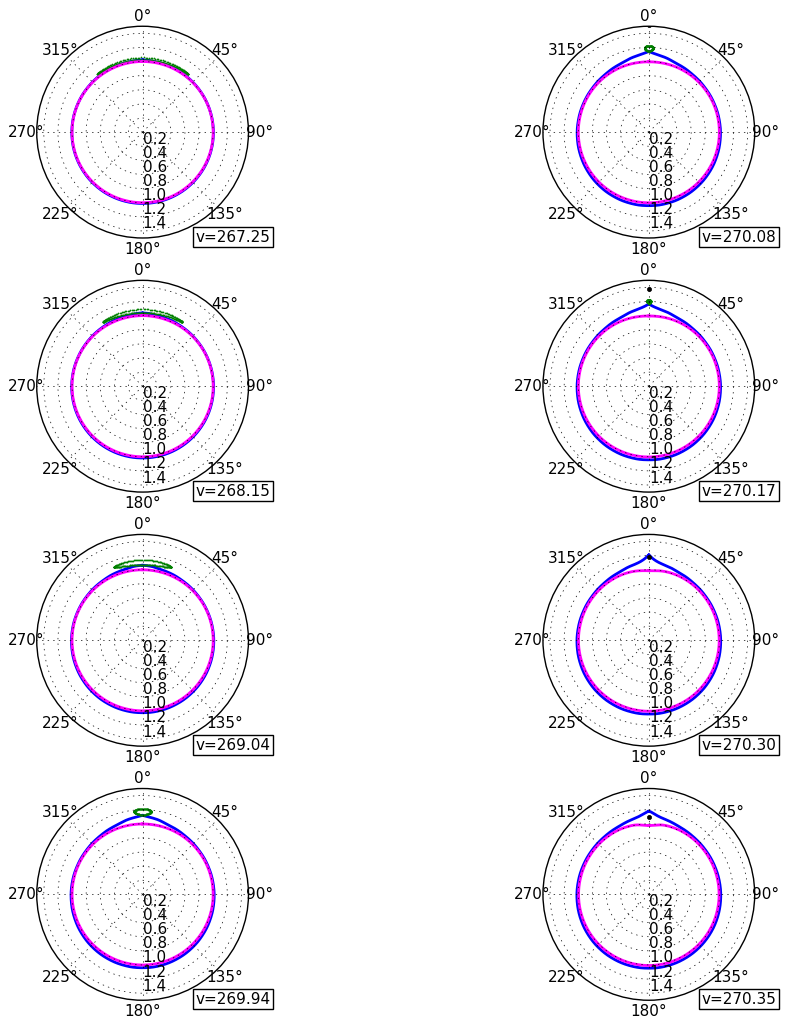}}
\caption{A visualization of the deformation of the coordinate shapes of the event and apparent horizons 
for $\mu/ M = 0.03$. Time advances from the top left down the columns. The green dots are null geodesics that join the event horizon forming caustics at $\theta=0$. The blue is the event horizon and the magenta is the apparent horizon.} \label{premerger1}
\end{figure}
\begin{figure}[h!]
\centerline{\includegraphics[scale=0.68]{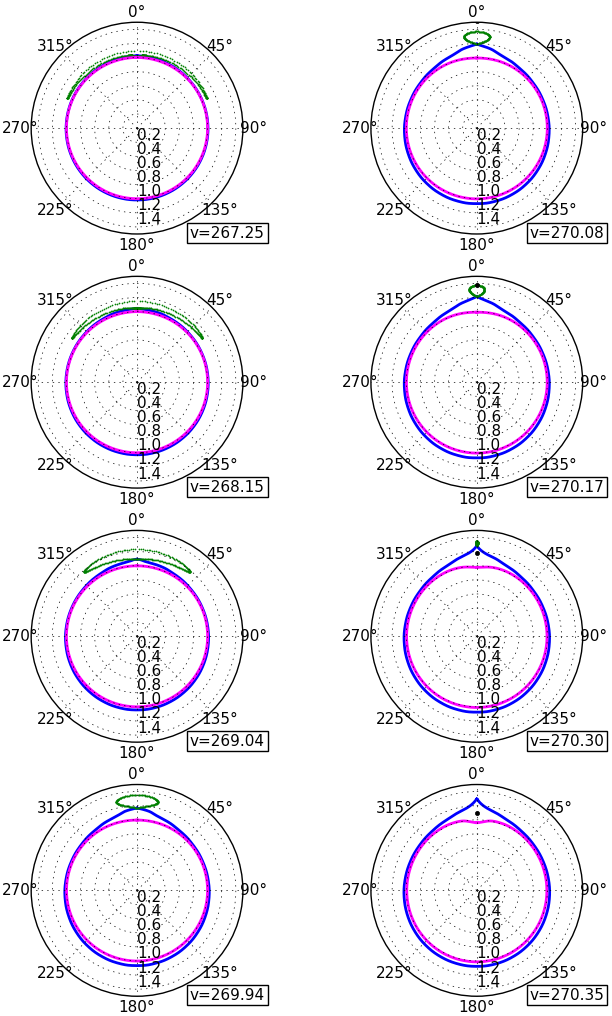}}
\caption{A visualization of the deformation of the coordinate shapes of the event and apparent horizons for $\mu/ M = 0.05$. Time advances from the 
top left down the columns. The green dots are null geodesics that join the event horizon forming caustics at $\theta=0$. The blue is the event horizon 
and the magenta is the apparent horizon. Note that the ``bubble'' of geodesics joining the event horizon disappears only after the particle crosses. 
This is an effect of the perturbation theory.    } \label{premerger2}
\end{figure}
\begin{figure}[h!]
\centerline{\includegraphics[scale=0.70]{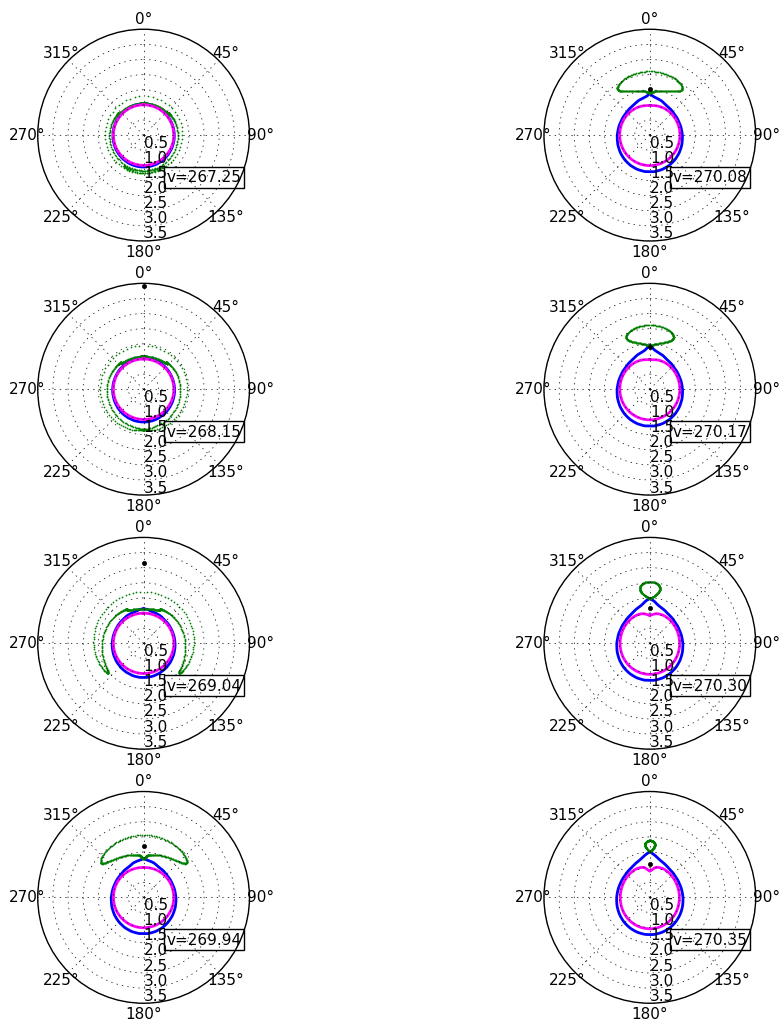}}
\caption{A visualization of the deformation of the coordinate shapes of the event and apparent horizons for $\mu/ M = 0.15$. Time advances from the top left down the columns. The green dots are null geodesics that join the event horizon forming caustics at $\theta=0$. The blue is the event horizon and the magenta is the apparent horizon. The perturbation theory can no longer to trusted for this ratio and so this case is included mainly to demonstrate effects arising from the failure.} \label{premerger3}
\end{figure}
\begin{figure}[h]
\centerline{\includegraphics[scale=0.666]{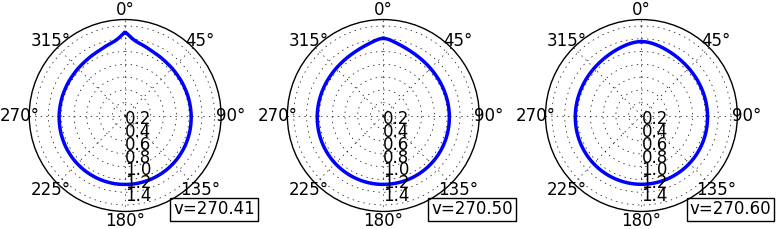}}
\caption{After the merger the black hole returns to its spherical shape, although slightly larger because of the contribution from the mass of the particle. This figure has $\mu/M = 0.05$ and so matches with FIG.~\ref{premerger2}. } \label{postmerger}
\end{figure}

Event horizons generated by these simulations are shown in FIG.~\ref{premerger1}--\ref{postmerger}. In each of these the mass of the large black 
hole is $M=1/2$ while the mass of the particle is $\mu = 0.015$ (FIG.~\ref{premerger1}), $\mu = 0.025$ (FIG.~\ref{premerger2},\ref{postmerger}) 
and $\mu = 0.075$ (FIG.~\ref{premerger3}). In the first three figures we see how the horizon deforms prior to $v_m$. The blue line is the event 
horizon and black dot illustrates the location of the particle. The green dots are null geodesics that are in the vicinity of the black hole. Given that
we are working in perturbation theory the smaller the ratio of $\mu/M$ the more accurate our results and so in particular FIG.~\ref{premerger1} ($\mu/M = 0.03$) should be the most accurate,  FIG.~\ref{premerger2} ($\mu/M = 0.05$) should still be pretty good but in FIG.~\ref{premerger3} ($\mu/M = 0.15$) is leaving the regime of applicability for the approximation. This last case is included mainly to show the effects of the approximation breaking
down. 

Then several interesting effects can be observed:
\begin{enumerate}[label=(\roman*)]
\item Before the merger there is a ``bubble'' of geodesics initially hugging close to the horizon that gradually moves away before joining the 
event horizon through caustics \cite{Siino:1997ix}. It appears that the last geodesic to join the horizon lies on the axis of symmetry in the top half of the bubble and the caustics end at the point when the particle 
crosses the event horizon. This is most clearly seen for $\mu/M = 0.03$. The disappearance of the caustics appears to be slightly post-merger 
for $\mu/M = 0.05$ however this is probably an effect of the perturbation theory beginning to fail as the lag is even more dramatic
 for $\mu/M = 0.15$. 
  
Similar behaviours for the ``bubble'' of geodesics destined to join the event horizon can be seen in full non-EMR numerical relativity \cite{Bohn:2016} 
as well as the exact Schwarzschild calculations of Emparan and Mart\'inez \cite{emparan_email}. 
 
\item As a consequence of these caustic points, the event horizon appears to be stretched towards the infalling particle. Similar effects can 
be seen in full numerical relativity \cite{Bohn:2016}, the exact Schwarzschild calculations   \cite{Emparan:2016} and the Hamerly-Chen
impulse approximation \cite{Hamerly:2010}. 

\item Post-merger (FIG.~\ref{postmerger}) the sharp caustic point rounds out and the event horizon returns to sphericity. Again similar effects
can be seen in the aforementioned studies. 
\end{enumerate}

Physically these effects are all a result of the focussing effects of the infalling particle. Because of its presence,  some near-horizon null geodesics 
that would have otherwise escaped are captured on the event horizon. The associated caustics occur  away from the original $r=2M$ surface in the 
direction of the particle. After it crosses the event horizon, the particle can no longer focus external geodesics. 

%
%
%

Before moving on it is worth reiterating the limitations of the approximations that we have made. First we are working in first
 order perturbation theory and so close to the particle, where fields become strong we cannot expect high accuracy. Then within the 
 perturbation theory the point particle is further approximated by the $l = 25$ cut-off. The effect of this 
is to replace the point particle with a shell of stress-energy of radius $r= r_p(t)$. As is discussed in Appendix \ref{ecviolations}, this is strongly 
peaked at $\theta =0$ but is also non-zero (and sometimes even energy-condition violating) at other $\theta$. 

Combined these approximations model the physical spacetime well away from the particle but poorly close to $r=r_p(t)$ and $\theta=0$. In particular
a real point mass would be masked by its own horizon and this is something that we cannot see in our model. Given that we do not have a horizon 
around the point mass we also cannot see any merger of horizons like that seen in  \cite{Emparan:2016}. 

That said we can have a high degree of confidence in features seen after the particle crosses the event horizon. 
Keep in mind that the event horizon is anchored in the future not the past: null rays are traced back from the future. As such once the particle is safely
inside the black hole the problems associated with the approximations similarly disappear. For small $\mu/M$ it is only in the approach where the 
various approximations can manifest themselves.

\subsection{Apparent Horizon}
Next we turn to the apparent horizon\footnote{Here we use apparent horizon in its colloquial sense as the outermost MOTS for the large black hole.} and see how it is deformed by the infalling particle. The method we use 
is a standard one \cite{Dennison:2006,baumgarte2010numerical} modified for small perturbations from the background geometry. We foliate the spacetime into spacelike slices, $\Sigma$, with the normal vector,
\begin{align}\label{slicevector}
\tilde{n}_\mu = \left( -\sqrt{f}+\frac{1}{2 \sqrt{f}} p_{tt} \right)\delta^t_\mu
\end{align}
which is the Schwarzschild time with a normalization $\tilde{n}^\mu \tilde{n}_\mu=-1$. Let $S$ be a closed two-dimensional surface in $\Sigma$ and let $\tilde{s}^\mu$ be the spacelike normal to $S$, hence, $\tilde{s}^{\mu} \tilde{s}_{\mu}=1$. For all points on $S$ we have outward and inward pointing null geodesics whose tangents, denoted $\tilde{k}^\mu$ and $\tilde{l}^\mu$ respectively, can be written as,
\begin{equation}
\tilde{k}^\mu = \frac{1}{\sqrt{2}} \left( \tilde n^\mu + s^{\mu} \right ) \hspace{0.5 cm} \text{and} \hspace{0.5 cm} \tilde{l}^\mu = \frac{1}{\sqrt{2}} \left( \tilde n^\mu - s^{\mu} \right )
\end{equation}
Then the metric, $\tilde{m}_{\mu\nu}$, on $S$ can be written as,  $\tilde{m}_{\mu\nu}= \tilde{g}_{\mu\nu} + \tilde{k}_\mu \tilde{l}_\nu + \tilde{l}_\mu \tilde{k}_\nu $. Further, the expansion of the outgoing null rays is given by, $\Theta= \tilde{m}^{\mu\nu} \tilde{\nabla}_{\mu } \tilde{ k_\nu}$. The apparent horizon is then defined to be a marginally outer trapped surface (MOTS) on which the expansion of the outward null geodesics vanishes, $\Theta=0$. 

We can define $S$ as a level surface of some scalar function $\tilde{\tau}(x^i)=0$, where the $x^i$ are coordinates on $\Sigma$. In the axisymmetric case, using Schwarzschild coordinates, we write, 
\begin{equation}
\tilde{\tau}(r,\theta)=r-\tilde{h}(\theta)
\end{equation}
and a normal and tangent to $S$ are
\begin{equation}
\tilde{m}_i = \partial_i \tilde{\tau} = (1,-\partial_{\theta} \tilde{h} ,0)\hspace{0.5 cm} \text{and} \hspace{0.5 cm}  \tilde{u}^i=\partial_\theta x^i = (\partial_\theta \tilde{h},1,0)
\end{equation}
Then the condition for vanishing expansion can be written as
\begin{equation}\label{apphorizon}
\partial_\theta^2 \tilde{h} = - (\tilde{X} + \tilde{s}^2 \tilde{Y}) - \frac{\tilde{s}}{\sqrt{\tilde{\gamma}^{(2)}}}(\tilde{P}+\tilde{s}^2 \tilde{Q})
\end{equation}
where,
\begin{align}
&\tilde{X}:= \tilde{\Gamma}^A_{B C} \tilde{m}_A \tilde{u}^B \tilde{u}^C,  &\tilde{Y}&:= \tilde{\gamma}^{\phi\phi} \tilde{\Gamma}^A_{\phi\phi} \tilde{m}_A ,\\
&\tilde{P}:= \tilde{K}_{AB} \tilde{u}^A \tilde{u}^B, &\tilde{Q}&:= \tilde{\gamma}^{\phi \phi} \tilde{K}_ {\phi \phi}, \\
&\tilde{s}^2:=\tilde{\gamma}_{AB} \tilde{u}^A\tilde{u}^B,
\end{align}
$\tilde{\gamma}_{\mu\nu}$ is the induced metric on $\Sigma$, the indices $A,B$ run over $(r,\theta)$, and $\tilde{\gamma}^{(2)}$ is the determinant of $\tilde{\gamma}_{AB}$. Since our approach is perturbative to linear order we will use the \emph{ansatz} $\tilde{h}(\theta)=2M+\delta h(\theta)$ and keep terms linear in $\delta$. Further evaluating $\psi(v)$ on $\mathcal{H}_1$, which allows us to switch to ingoing coordinates, (\ref{apphorizon}) becomes,
\begin{equation}\label{apphorzeroexp}
\left( \partial_\theta^2 + \cot \theta \partial_\theta 	\right) \delta h -2 \delta h=- 2\left[ \partial_v^2 \psi^l - \left(\lambda+2\right) \partial_v\psi^l \right] Y^l.
\end{equation}
Notice that $\left( \partial_\theta^2 + \cot \theta \partial_\theta 	\right)$ is the Laplacian on the unit sphere. To solve (\ref{apphorzeroexp}) we assume that $\delta h$ can be expanded in 
($m=0$) spherical harmonics,
\begin{equation}\label{apphora1}
\delta h(\theta) = \sum_l a^l Y^{l0} (\theta),
\end{equation}
inserting this into (\ref{apphorzeroexp}) we find,
\begin{equation}\label{apphora2}
a^l= \lambda^{-1} \partial_v^2 \psi^l - \partial_v \psi^l.
\end{equation}
Inserting (\ref{apphora2}) into (\ref{apphora1}) we can construct the deformation of the apparent horizon. This is shown as the magenta curve in FIG.~
\ref{premerger1}-\ref{premerger3}. For early times, when the particle is far from the black hole the apparent horizon coincides with $r=2M$. As for the
event horizon the effects of the approaching particle are concentrated on the piece of the horizon facing the particle. However the apparent 
horizon behaves quite differently from the event horizon. While the event horizon puckered out towards the particle the apparent horizon recedes
(first seen about $v=270.30$ in all simulations) and appears to become concave outwards as the particle comes closer.

%

Intuitively  this behaviour can be understood in the following way. As the particle approaches the black hole, it bends light rays towards it. In particular this increases the outward expansion of 
congruences of null geodesics inside the $r=2M$ surface and thereby increases their (initially negative) expansion. Thus the new 
 $\Theta=0$ surface is inside the old $r=2M$ surface.  This recession does not violate the rule that apparent horizons cannot decrease in area \cite{Hayward:1993wb,Ashtekar:2004cn,Andersson:2005gq,Bousso:2015qqa}: our approximation is only first order and area change is a second order effect. To the order of accuracy of our approximation the area can't change.
%

\ivc{There is a gauge issue to consider here. Apparent horizons are creatures of their particular foliation and in general the time-evolved apparent horizon of one foliation will not match that 
from a different slicing of spacetime. Hence a proper physical understanding of the recession would include an investigation of its foliation dependence. That is beyond the scope of the current
paper but we note that this is not the first time that such an effect has been observed.}
%
%
%
%
%
\ivc{ A similar effect was seen in the analysis of the MOTSs in initial data for head-on collisions in a non-perturbative setting \cite{Mosta:2015}\footnote{Referring back to the concerns of the
 previous paragraph, in that paper the total horizon area was seen to increase in spite of the recession from the smaller black hole.}. Hence we expect this recession to be robust for reasonable
choices of foliation. 
}


{
Finally we note that $v=270.35$ is probably as far as we can expect 
our linear perturbation theory to return reasonable results for the apparent horizon. 
For example with ${\mu}/{M} = 0.05$, by  $v=270.35$ the
apparent horizon of the small black hole (that we are modelling as a particle) is approaching that of the large black hole. Hence we are now in a region where the 
gravitational field that we have assumed to be small is becoming large. \ivc{Indeed just after this point, our horizon finding methods fail with the apparent horizon diverging 
through and away from the event horizon. } 

\ivc{This undesirable (and physically impossible!) behaviour results from trying to push the horizon through an area where our approximations are no longer good:
the perturbations are no longer small and we are probing at a resolution where the $l=25$ cut-off becomes problematic. 
As was emphasized in \cite{Booth:2017fob}, identifying the location of an apparent horizon strongly depends on local features of the geometry: there are many surfaces of vanishing null expansion running through all points of a spacetime and the only thing distinguishing a MOTS is that it is a closed surface. }
As such, once we lose control of those details in a region through which the horizon should pass, we can no longer locate it. 

\ivc{In our case, at $v=270.35$ the apparent horizon should close near the infalling particle. However our description of the geometry is not good in that region and 
so the closure does not happen in the expected way. }
By contrast the highly non-local event horizon calculations are more robust. Problems in one small region only affect a few geodesics 
and so the global evolution is not seriously perturbed.


\section{Discussion }\label{conclusion}
In this paper we modelled an EMR merger by using perturbation techniques for Schwarzschild black holes. Specifically, we used the stress tensor of a point particle to model the gravitational field of a small black hole/point particle and used Zerilli's formalism to solve the linearized Einstein's equations. We then studied 
the deformation of the event and apparent horizons as the particle approaches the black hole  (FIGS. \ref{premerger1}--\ref{postmerger}). 

Given the teleological nature of event horizons we used final post-merger boundary conditions and integrated backwards in time to get the full horizon. 
To ensure a smooth horizon evolution we fixed constants in (\ref{constant1}) to (\ref{constant3}), whereby we see that the discontinuity in the initial data plays a major role in the deformation of the 
event horizon. This discontinuity arises from doing a multipole expansion of the Brill-Lindquist type initial data \cite{Lousto:1996}. Such a dependence on initial data for perturbations is lost when 
analysing solutions of (\ref{master}) in the frequency domain. We also saw how there is a ``bubble'' of null rays that emerges between the particle and the black hole just before the merger, shown in 
green in FIG. \ref{premerger1}--\ref{premerger2}. These are null rays that join the event horizon at the top of the black hole by crossing each other and forming caustics. We noted that this bubble of null rays and associated
caustics are a standard feature of black hole mergers and commonly seen in full non-linear treatments (for example \cite{Bohn:2016}).  

For locating the apparent horizon we used a standard approach \citep{baumgarte2010numerical} simplified to the case of linear perturbations. Perhaps the most interesting observation was that 
as the particle makes its final approach to the black hole, the region of the apparent horizon facing the particle appears to  recede away from it, ultimately becoming concave outwards in 
the diagrams. This apparently
non-intuitive behaviour can be understood as resulting from null rays in the immediate vicinity of the unperturbed apparent horizon being attracted towards particle. Hence relative to the unperturbed
metric the apparent horizon moves inwards. At the same time we noted that the recession doesn't result in a decrease in the area of the horizon: in our first order expansion the area is invariant.


It is interesting to compare the quite different evolutions of the  event versus apparent horizons. The event horizon is ``attracted'' towards the particle as one might naively expect, while the 
apparent horizon initially recedes. Further during the final approach the event horizon puckers to form a caustic cusp as new null generators join the horizon while the apparent horizon remains 
smooth as long as it can be found


The appearance of caustics is not a consequence of our point source. 
For one we have noted that our cut-off of higher $l$ modes ``smears out'' the particle and so despite our initial set-up we are not really dealing with a point source. However more generally caustics are a result of geodesic focussing and this does not require a point source. For example they are also seen in  \cite{Emparan:2016} where the smaller black hole has a Schwarzschild geometry and is not a point source, as well as in the full non-linear treatment of mergers with arbitrary initial parameters \cite{Bohn:2016,Matzner941}. 

{The event horizon grows during the approach (thanks to its final boundary 
conditions) while the apparent horizon does not (thanks to the linear approximations). However this is  realistic. During this physical process most of the mass change in the large
black hole comes directly from the infalling particle. The event horizon grows ``in anticipation'' of the absorption however the apparent horizon should not substantially change in size until the particle actually is absorbed.} 

Overall we have found that for an EMR merger many key features can be captured by using standard perturbation methods. However some features escape us. Specifically any features related to the event and apparent horizon of the smaller black hole are lost, simply because we use a point particle approximation. Perhaps the matched asymptotic methods suggested in \cite{Emparan:2016} could be used to focus on the event horizon geometry of the smaller black hole. Further the appearance of a common apparent horizon \cite{Bishop1982, Cadez1973, Anninos:1994ay, Cook:1992} that is usually seen in arbitrary mass ratio mergers is also lost. This feature in principle could also be recovered by upgrading the particle to a Schwarzschild geometry rather than a point stress tensor. In future papers we intend to return to this issue as well as considering more general interactions than the head-on collisions studied here. 

\acknowledgements
{ IB thanks Abhay Ashtekar and Aaron Zimmerman for a conversation at the 15th Canadian Conference on General Relativity and Relativistic Astrophysics which inspired this paper. UH would like to thank Hari Kunduri, Harald Pfeiffer  and Aaron Zimmerman for their insights and suggestions. 
We would also like to thank Roberto Emparan for an enlightening correspondence following the posting of the first version of this paper to the arXiv 
which significantly influenced later versions \ivc{and Eric Poisson for his valuable comments contributed as part of UH's PhD defence}. IB is supported by NSERC Discovery Grant 261429-2013. 
During the course of this research UH was partially funded by a fellowship from the School of Graduate Studies at Memorial University as well as stipends from NSERC Discovery Grants 261429-2013 
and 
418537-2012.

\appendix
\section{Choices for the $l=1$ gauge}\label{l1pert}
The $l=1$ perturbation is given by,
\begin{align}
H_0^{10}&=\frac{\left[f_0(t)+(r^3/M) \ddot{f}_0(t)\right]}{3(r-2M)^2} \Theta(r-r_p(t))\\
H_1^{10}&=- \frac{r \dot{f}_0(t)}{(r-2M)^2} \Theta(r-r_p(t)) \\
H_2^{10}&=\frac{f_0 (t)}{(r-2M)^2} \Theta(r-r_p(t))
\end{align}
where $r_p(t)$ is the trajectory, $\Theta$ is the Heaviside function and,
\begin{equation}
f_0(t)=8 \pi \mu (r_p(t)-2M) \sqrt{3/4\pi}.
\end{equation}
 Clearly, this is a gauge where the perturbation is non-zero outside a sphere of radius $r=r_p(t)$. As mentioned above, when computing the event horizon deformation we need to evaluate the right hand side of Eqs. (\ref{deltathetav}) and (\ref{deltauv}) at $r=2M$. We find,
 \begin{equation}\label{delgvv}
\delta g_{VV} = \frac{1}{4 \kappa^2} \frac{1}{V^2} \frac{1}{r(r-2M)} \left[ \frac{4}{3} f_0(v) - 2r\partial_v f(v)+ \frac{r^3}{3M} \partial^2_v f_0(v)\right] Y^1
\end{equation}
 where $f_0(v)\sim r_p(v)$ where $r_p(v)$ is the trajectory of the particle in Eddington-Finkelstein coordinates (see for example \cite{muller2008falling}). The term in the square brackets is finite and hence the whole quantity is undefined at $r=2M$. We can utilize the gauge transformation used by Zerilli in \cite{Zerilli:1971} to switch to a gauge where the perturbation is non-zero inside a sphere of radius $r=r_p(t)$. We find in this gauge,
 \begin{align}\label{rltrp}
    \delta g_{VV}=\frac{1}{4 \kappa^2} \frac{1}{V^2} \frac{1}{r(r-2M)}\left[-\frac{4}{3} f_0(t) + 2 \dot{f}_0(t) -\frac{r^3}{M} \ddot{f}_0(t) \right] Y^1    
    \end{align}  
where the dots denote derivatives w.r.t. $t$. Again, we see a similar situation as above where the term in the square brackets is finite and the whole quantity is undefined at $r=2M$. 

It is possible to go into a gauge where the perturbation is concentrated at the location of the particle in the form of a delta function. This gauge can be achieved by appropriately adding a Heaviside function to Zerilli's gauge transformation. In this gauge we have,
\begin{align}
\nonumber \delta g_{VV} =&\frac{1}{4 \kappa^2} \frac{1}{V^2} \bigg[ 4 \mu  \frac{r^2}{r_p(t)^3}  \frac{(r_p(t) - 2M)^2}{r-2M} + \frac{2r\mu}{M} \sqrt{\frac{2M}{r_p(t)}} \left( 1 - \frac{2M}{r_p(t)} \right) \left(1 + \frac{r_p(t)-2M}{r-2M}  \right) \\ &  -\frac{\mu}{Mr} (r-2M) (r_p(t) - 2M) \bigg] \delta(r-r_p(t))\sqrt{\frac{4\pi}{3}} Y^1.
\end{align}
We get the contribution from the delta function when we integrate in Eq. (\ref{deltauv}) but notice that the quantity in square brackets in front of the delta function vanishes when $r_p(t) \rightarrow 2M$ and $r \rightarrow 2M$. This nulls the effect of the delta function. Choosing the trivial solution for Eq. (\ref{deltauv}) we see that in this gauge the $l=1$ perturbation does not effect the horizon deformation. The same reasoning applies to Eq (\ref{deltathetav}) as $\delta \Gamma^\theta_{VV} \sim \partial_{\theta} \delta g_{VV}$. Further, since we compute the apparent horizon before the particle touches $r=2M$, the apparent horizon is not affected. 

\section{Energy condition violations and $l$ cut-off}\label{ecviolations}
In this appendix we demonstrate that the introduction of an $l=25$ cut-off in the spherical harmonic expansion of the wave-form is physically 
equivalent to the introduction of a spherical matter field at $r=r_p(t)$. For $l=25$ the energy density is strongly peaked at $\theta=0$ however it is 
not a perfect point source. While we do not believe that this approximation significantly affects our simulation it is useful to keep in mind when thinking
about the results of the simulations.





The stress tensor associated with a point particle is given by,
\begin{equation}
T^{ab}=\mu \int u^a u^b \delta(x^\nu - x_p^{\nu}(\tau)) (-g)^{-1/2} d \tau
\end{equation}
where $\{a,b\}=\{r,t\}$, $u^a$ is the four velocity of the particle and $x_p^{\nu}$ is the position of the particle. This  can be expanded in ($m=0$) spherical harmonics,
\begin{equation}
T^{ab}=\sum_{l=0}^{\infty} T^{ab}_{(l)} Y_l(\theta)
\end{equation}
where 
\begin{equation}
T^{ab}_{(l)} = \int T^{ab} Y^*_{l}(\theta') d \Omega'^2.
\end{equation}
\begin{figure}[h!]
\centering
\includegraphics[scale=0.55]{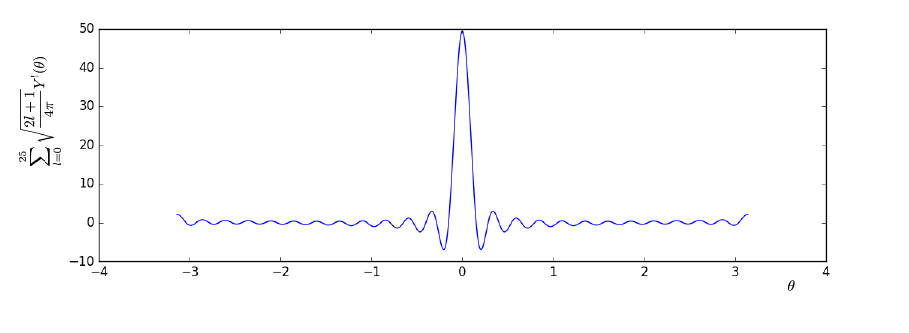}
\caption{A plot of sum in \cref{seriescut} with a cut off at $l=25$, the $x$-axis is $\theta$ in radians. Notice how there are regions where the sum is negative. }\label{cutofffig}
\end{figure} 

Further, it can be shown that,
\begin{equation}
\int T^{ab} Y^*_{l}(\theta') d \Omega'^2= \frac{8\pi}{r^2} \frac{u^a u^b}{u^t}  \delta(r-r_p(t)) Y_l(\theta_p(t)).
\end{equation}
Now, since the particle falls along the $z$-axis, its $\theta$ coordinate does not change: $\theta_p=0$. This implies that $ Y_l(0)=\sqrt{\frac{2 l +1}{4 \pi}} P_l (\cos(0))=\sqrt{\frac{2 l +1}{4 \pi}}$  for all  $t$. We have then,
\begin{equation}\label{seriescut}
T^{tt}=\mu \frac{\tilde{E}}{r^2f(r)} \delta(r-r_p(t)) \sum_{l=0}^\infty \sqrt{\frac{2 l +1}{4 \pi}} Y^l(\theta).
\end{equation}

If a finite cut-off is introduced in the $l$ series in \cref{seriescut} at $l_{\text{cut}}=25$, we get dependence on $\theta$ as shown in  \cref{cutofffig}. Notice that there are regions where the sum is negative, this implies that there are regions where $T^{tt}_{\text{cut}} <0$. This means that there are small energy condition violations associated with introducing a finite $l$ cut off. 

Such energy condition violations can, for example, cause the apparent horizon to be outside the event horizon in a dynamical setting (see,  for example,  \cite{Scheel:1994yn}). However they are very weak in our example and so relatively innocuous. 

\section{Numerical convergence and waves at infinity}\label{convergence}
To test our implementation of the numerical algorithm we make comparisons with the results of \cite{Martel:2001}.  Specifically, we reproduce the result presented in FIG. 3 of \cite{Martel:2001} where the particle falls in from rest with a starting position of $r_0/2M=40$ and the $l=2$ master function is evaluated at null infinity, $\mathcal{H}_2$. This is shown in FIG. \ref{comparetopoisson}, where it can be seen that we get the same result (for $\alpha=1$), we get identical results for other values $l$ and $r_0$ also.
\begin{figure}
\includegraphics[scale=1]{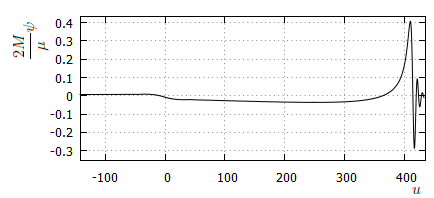}
\caption{The master function with $l=2$ evaluated at $\mathcal{H}_2$ with $r_0/2M=40$. Note that this is  virtually identical to FIG. 3 in \cite{Martel:2001} with $\alpha=1$.} \label{comparetopoisson}
\end{figure} 

We also demonstrate quadratic convergence of the code in an identical manner to the Appendix of 
\cite{Martel:2001}
by computing $\delta \psi$, as defined in the Appendix of 
\cite{Martel:2001}. 
For completeness the definition of $\delta \psi$ is reviewed here. The numerical algorithm is constructed to be convergent at second order, hence we expect,
\begin{equation}
\psi_N(\Delta) = \psi_{\text{exact}} + \Delta^2 \rho
\end{equation}
where $\psi_N$ is the numerical solution, $\psi_{\text{exact}}$ is the exact solution, $\Delta$ is step size and $\rho$ is an error function that is independent of $\Delta$. We may then define,
\begin{equation}
\delta \psi(\Delta) = \psi_N(2 \Delta) - \psi_N(\Delta).
\end{equation}
\begin{figure}[h]
\includegraphics[scale=1]{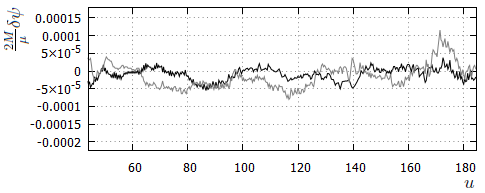}
\caption{Plot of $\delta \psi$ vs. $u$. The black curve is $\delta \psi$ with a resolution $\Delta=0.04$ and the grey curve is $4\delta \psi$ with a resolution of $\Delta = 0.02$. The agreement between these curves indicates quadratic convergence towards the exact solution. } \label{noisy}
\end{figure}
Notice that we have $\delta \psi(n \Delta) = n^2 \delta \psi (\Delta)$. We calculate $\delta \psi(\Delta)$ at a resolution of $\Delta = 0.02$ and $\Delta=0.04$. In FIG. \ref{noisy}   the black curve is $\delta \psi$ with a resolution $\Delta=0.04$ and the grey curve is $4\delta \psi$ with a resolution of $\Delta = 0.02$, the agreement of these curves indicates quadratic convergence.
\bibliography{horizon}

\begin{thebibliography}{10}

\bibitem{Abbott:2016}
B.~P. Abbott {\em et~al.}, ``{Observation of Gravitational Waves from a Binary
  Black Hole Merger},'' {\em Phys. Rev. Lett.}, vol.~116, no.~6, p.~061102,
  2016.

\bibitem{Zerilli:1971}
F.~J. Zerilli, ``{Gravitational field of a particle falling in a schwarzschild
  geometry analyzed in tensor harmonics},'' {\em Phys. Rev.}, vol.~D2,
  pp.~2141--2160, 1970.

\bibitem{Martel:2001}
K.~Martel and E.~Poisson, ``{A One parameter family of time symmetric initial
  data for the radial infall of a particle into a Schwarzschild black hole},''
  {\em Phys. Rev.}, vol.~D66, p.~084001, 2002.

\bibitem{Spallicci:2010}
A.~Spallicci, ``{Free fall and self-force: An Historical perspective},'' {\em
  Fundam. Theor. Phys.}, vol.~162, pp.~561--603, 2011.
\newblock [,561(2010)].

\bibitem{Lousto:1997}
C.~O. Lousto and R.~H. Price, ``{Understanding initial data for black hole
  collisions},'' {\em Phys. Rev.}, vol.~D56, pp.~6439--6457, 1997.

\bibitem{Emparan:2016}
R.~Emparan and M.~Martinez, ``{Exact Event Horizon of a Black Hole Merger},''
  {\em Class. Quant. Grav.}, vol.~33, no.~15, p.~155003, 2016.

\bibitem{Hamerly:2010}
R.~Hamerly and Y.~Chen, ``{Event Horizon Deformations in Extreme Mass-Ratio
  Black Hole Mergers},'' {\em Phys. Rev.}, vol.~D84, p.~124015, 2011.

\bibitem{Mosta:2015}
P.~M{\"o}sta, L.~Andersson, J.~Metzger, B.~Szil{\'a}gyi, and J.~Winicour, ``The
  merger of small and large black holes,'' {\em Classical and Quantum Gravity},
  vol.~32, no.~23, p.~235003, 2015.

\bibitem{Moncrief:1974}
V.~Moncrief, ``{Gravitational perturbations of spherically symmetric systems.
  I. The exterior problem.},'' {\em Annals Phys.}, vol.~88, pp.~323--342, 1974.

\bibitem{Brill:1963}
D.~R. Brill and R.~W. Lindquist, ``{Interaction energy in geometrostatics},''
  {\em Phys. Rev.}, vol.~131, pp.~471--476, 1963.

\bibitem{Lousto:1996}
C.~O. Lousto and R.~H. Price, ``{Headon collisions of black holes: The Particle
  limit},'' {\em Phys. Rev.}, vol.~D55, pp.~2124--2138, 1997.

\bibitem{Anninos:1994}
P.~Anninos, D.~Bernstein, S.~Brandt, J.~Libson, J.~Masso, E.~Seidel, L.~Smarr,
  W.-M. Suen, and P.~Walker, ``{Dynamics of apparent and event horizons},''
  {\em Phys. Rev. Lett.}, vol.~74, pp.~630--633, 1995.

\bibitem{detweiler2004low}
S.~Detweiler and E.~Poisson, ``Low multipole contributions to the gravitational
  self-force,'' {\em Physical Review D}, vol.~69, no.~8, p.~084019, 2004.

\bibitem{Siino:1997ix}
M.~Siino, ``{Topology of event horizon},'' {\em Phys. Rev.}, vol.~D58,
  p.~104016, 1998.

\bibitem{Bohn:2016}
A.~Bohn, L.~E. Kidder, and S.~A. Teukolsky, ``{Toroidal Horizons in Binary
  Black Hole Mergers},'' {\em Phys. Rev.}, vol.~D94, no.~6, p.~064009, 2016.

\bibitem{emparan_email}
R.~Emparan , Email exchange, May 2017.

\bibitem{Dennison:2006}
K.~A. Dennison, T.~W. Baumgarte, and H.~P. Pfeiffer, ``{Approximate initial
  data for binary black holes},'' {\em Phys. Rev.}, vol.~D74, p.~064016, 2006.

\bibitem{baumgarte2010numerical}
T.~Baumgarte and S.~Shapiro, {\em Numerical Relativity: Solving Einstein's
  Equations on the Computer}.
\newblock Cambridge University Press, 2010.

\bibitem{Hayward:1993wb}
S.~Hayward, ``{General laws of black hole dynamics},'' {\em Phys.Rev.},
  vol.~D49, pp.~6467--6474, 1994.

\bibitem{Ashtekar:2004cn}
A.~Ashtekar and B.~Krishnan, ``{Isolated and dynamical horizons and their
  applications},'' {\em Living Rev.Rel.}, vol.~7, p.~10, 2004.

\bibitem{Andersson:2005gq}
L.~Andersson, M.~Mars, and W.~Simon, ``{Local existence of dynamical and
  trapping horizons},'' {\em Phys.Rev.Lett.}, vol.~95, p.~111102, 2005.

\bibitem{Bousso:2015qqa}
R.~Bousso and N.~Engelhardt, ``{Proof of a New Area Law in General
  Relativity},'' {\em Phys. Rev.}, vol.~D92, no.~4, p.~044031, 2015.

\bibitem{Booth:2017fob}
I.~Booth, H.~K. Kunduri, and A.~O'Grady, ``{Unstable marginally outer trapped
  surfaces in static spherically symmetric spacetimes},'' {\em Phys. Rev.},
  vol.~D96, no.~2, p.~024059, 2017.

\bibitem{Matzner941}
R.~A. Matzner, H.~E. Seidel, S.~L. Shapiro, L.~Smarr, W.-M. Suen, S.~A.
  Teukolsky, and J.~Winicour, ``Geometry of a black hole collision,'' {\em
  Science}, vol.~270, no.~5238, pp.~941--947, 1995.

\bibitem{Bishop1982}
N.~T. Bishop, ``The closed trapped region and the apparent horizon of two
  schwarzschild black holes,'' {\em General Relativity and Gravitation},
  vol.~14, no.~9, pp.~717--723, 1982.

\bibitem{Cadez1973}
A.~\v{C}ade\v{z}, ``Apparent horizons in the two-black-hole problem,'' {\em
  Annals of Physics}, vol.~83, no.~2, pp.~449 -- 457, 1974.

\bibitem{Anninos:1994ay}
P.~Anninos, D.~Bernstein, S.~Brandt, J.~Libson, J.~Masso, E.~Seidel, L.~Smarr,
  W.-M. Suen, and P.~Walker, ``{Dynamics of apparent and event horizons},''
  {\em Phys. Rev. Lett.}, vol.~74, pp.~630--633, 1995.

\bibitem{Cook:1992}
G.~B. Cook and A.~M. Abrahams, ``{Horizon structure of initial data sets for
  axisymmetric two black hole collisions},'' {\em Phys. Rev.}, vol.~D46,
  pp.~702--713, 1992.

\bibitem{muller2008falling}
T.~M{\"u}ller, ``Falling into a schwarzschild black hole,'' {\em General
  Relativity and Gravitation}, vol.~40, no.~10, pp.~2185--2199, 2008.

\bibitem{Scheel:1994yn}
M.~A. Scheel, S.~L. Shapiro, and S.~A. Teukolsky, ``{Collapse to black holes in
  Brans-Dicke theory. 2. Comparison with general relativity},'' {\em Phys.
  Rev.}, vol.~D51, pp.~4236--4249, 1995.

\end{thebibliography}
\bibliographystyle{ieeetr}
\end{document}